\documentclass[aps,twocolumn]{revtex4}
\usepackage[spanish]{babel}
\usepackage[latin1]{inputenc}
 \usepackage{graphicx}
 \usepackage{amssymb}
 \usepackage{amsfonts}
\usepackage{amsmath}

\begin{document}

\title{¿Los \'atomos como osciladores perfectos?}

\author{Gabriel J. Gil P\'erez$^{a}$ y Augusto Gonz\'alez$^{b}$}

\affiliation{$^{a}$Facultad de F\'isica, Universidad de La Habana, 
 CP 10400, Ciudad de La Habana, Cuba\\
$^{b}$Instituto de Cibern\'etica, Matem\'atica y F\'isica, CP 10400,
 Ciudad de La Habana, Cuba}

\begin{abstract}
By using Supersymmetric Quantum Mechanics and Semiclassical
Quantization, one may argue that the low-lying excited states of any quantum 
system can be modeled by a set of harmonic oscillators. In the present
paper, we fit the experimental excitation spectra of atoms with atomic
number $2\le Z\le 36$ to a simple harmonic oscillator model with two
parameters: the number of degrees of freedom, $d$, and the effective
frequency, $\omega$. The obtained $\hbar\omega$ takes values around 
0.03 (in atomic units), whereas $d$ shows clear shell filling effects, 
that is, takes high values for the noble gases, suggesting collective 
oscillations of the electrons occupying the last shell.
\end{abstract}

\maketitle

\section{Introducción}

El presente trabajo tiene la intenci\'on de ser un modesto homenaje al 
Prof. Marcos Moshinsky, 
fallecido recientemente, quien fue un defensor del papel jugado por el
oscilador arm\'onico en la F\'isica moderna y un pionero en la
construcci\'on de bases de funciones de tipo oscilador para 
la descripci\'on mecanocu\'antica de los n\'ucleos at\'omicos
\cite{Moshinsky}.

En contraste con el rigor matem\'atico del Prof. Moshinsky, vamos a
seguir un estilo muy cualitativo e intentar, con argumentos simples y
ajustes a datos experimentales, dar respuesta a la interrogante de si el
espectro de excitaciones de los \'atomos se puede modelar con
osciladores arm\'onicos. Esta interrogante se puede desdoblar, en
realidad, en dos aspectos muy interesantes, que abordaremos
someramente.

Por un lado, es bastante extendida en la comunidad de f\'isicos la 
intuici\'on de que cualquiera sea el sistema cu\'antico o el potencial de
interacci\'on entre part\'iculas, el espectro de excitaciones a
bajas energ\'ias es siempre el de un sistema de osciladores. 
Daremos un argumento a favor de este cuadro cualitativo apelando a 
la cuantizaci\'on semicl\'asica en un potencial efectivo sugerido por la 
denominada Mec\'anica Cu\'antica Supersim\'etrica. Este argumento
fue formulado en [\onlinecite{mios}] y aplicado a sistemas
modelos de tres part\'iculas.

\begin{figure}[ht]
  \includegraphics[width=.5\linewidth,angle=0]{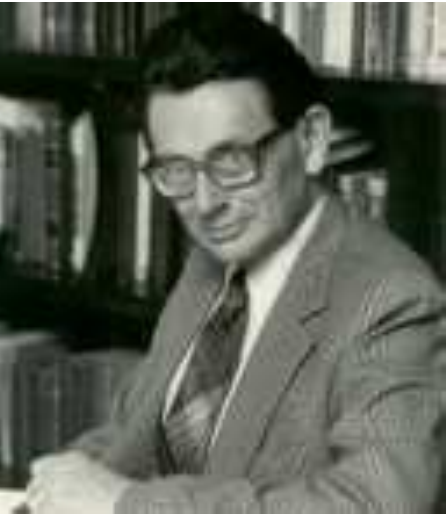}
  \caption{El Prof. Marcos Moshinsky, premio Pr\'incipe de Asturias de
  1988 y Premio UNESCO de las Ciencias en 1997.}
\end{figure}

Por otro lado, es tambi\'en muy atractiva la idea de ajustar con pocos 
par\'ametros efectivos datos experimentales o calculados de una magnitud
f\'isica. Esto es
muy com\'un en el quehacer de la F\'isica experimental. Como ejemplo en
el \'ambito de la teor\'ia de mol\'eculas, muy cercano al esp\'iritu del
presente trabajo, podemos citar el hecho
conocido de que el espectro de oscilaciones de las prote\'inas puede 
ajustarse muy bien con una sola constante de fuerza para describir la
interacci\'on entre cualquier par de \'atomos \cite{Tirion}. En nuestro
trabajo utilizaremos la excelente compilaci\'on del NIST sobre los
espectros de excitaciones de \'atomos \cite{NIST} y ajustaremos cada
espectro con s\'olo dos par\'ametros: la frecuencia ($\omega$) y la 
dimensi\'on efectiva ($d$) del oscilador. Para un \'atomo como el Na, 
por ejemplo, con un
solo electr\'on de valencia, uno esperar\'ia $d\approx 3$ y $\omega$
describir\'ia la interacci\'on de ese electr\'on con el resto de los
electrones internos. En el caso m\'as general, $d/3$ ser\'ia un indicador 
del n\'umero efectivo de electrones que participan en las oscilaciones.

\subsection{El cuadro de osciladores para los estados excitados}

Consideremos un sistema cu\'antico de varias part\'iculas, descrito por
el Hamiltoniano:

\begin{equation}\label{hamiltoniano}
H= \sum_{i}\left\{ \frac{p^{2}_{i}}{2 \mu_{i}} + V\right\},
\end{equation}

\noindent
donde $V$ da cuenta de la interacci\'on. Si la función de onda
del estado base del sistema, $\psi_{0}= \exp-W/\hbar$ y la
energ\'ia, $E_{0}$, son conocidas, podemos escribir la siguiente
expresi\'on para el potencial:

\begin{equation}
V - E_{0}=\sum_{i}\frac{1}{2\mu_i}\left\{(\nabla_i W)^2 - 
 \hbar (\Delta_i W)\right\}.
\end{equation}

\noindent
Esta expresi\'on se conoce como ``representaci\'on del estado base'' y
tiene su origen en la denominada ``Mec\'anica Cu\'antica
Supersim\'etrica'' \cite{susyqm}, la cual se construye por analog\'ia
con la supersimetr\'ia en la F\'isica de part\'iculas, pero en realidad
trata sobre relaciones de casi isoespectralidad entre hamiltonianos. El
denominado ``superpotencial'':

\begin{equation}
U=\sum_{i}\frac{1}{2\mu_i}(\nabla_i W)^2,
\end{equation}

\noindent
tiene las siguientes propiedades:

\begin{itemize}
\item[-] $U\ge 0$ a\'un cuando $V$ no sea acotado por debajo.
\item[-] El mínimo, $U=0$, corresponde a $\nabla_i W=0$, es decir la
configuraci\'on de máxima probabilidad. Esto es una extensi\'on del
concepto de radio de Bohr a sistemas con muchos grados de libertad.
\item[-] $U$ difiere de $V - E_{0}$ en términos que son formalmente
de orden superior en ${\hbar}$. Lo que sugiere que la
regla de cuantización de Einstein-Keller-Maslov:

\begin{equation}
\oint{\sum_{i}\vec p_i\cdot\vec dr_i}={(n_{\alpha}+
\frac{1}{4}m_{\alpha})}{\hbar}
\label{ec4}
\end{equation}

podr\'ia aplicarse en $U$. 
\item[-] El estado base, al que corresponde energ\'ia de excitaci\'on 
igual a cero, es descrito exactamente por la Ec. (\ref{ec4}) con 
\'indices $n_\alpha=0$, $m_\alpha=0$, por eso dicha condici\'on de 
cuantizaci\'on es buena no s\'olo cuando  $n_{\alpha}$ es grande (como 
sucede con toda semicl\'asica), sino tambi\'en cuando los $n_\alpha$ 
son peque\~nos.
\item[-] La Ec. (\ref{ec4}) con $m_\alpha=0$ es exacta en toda una clase
de problemas de una part\'icula en una dimensi\'on, que incluye al
potencial de Coulomb.
\end{itemize}

\begin{figure}[ht]
  \includegraphics[width=.8\linewidth,angle=0]{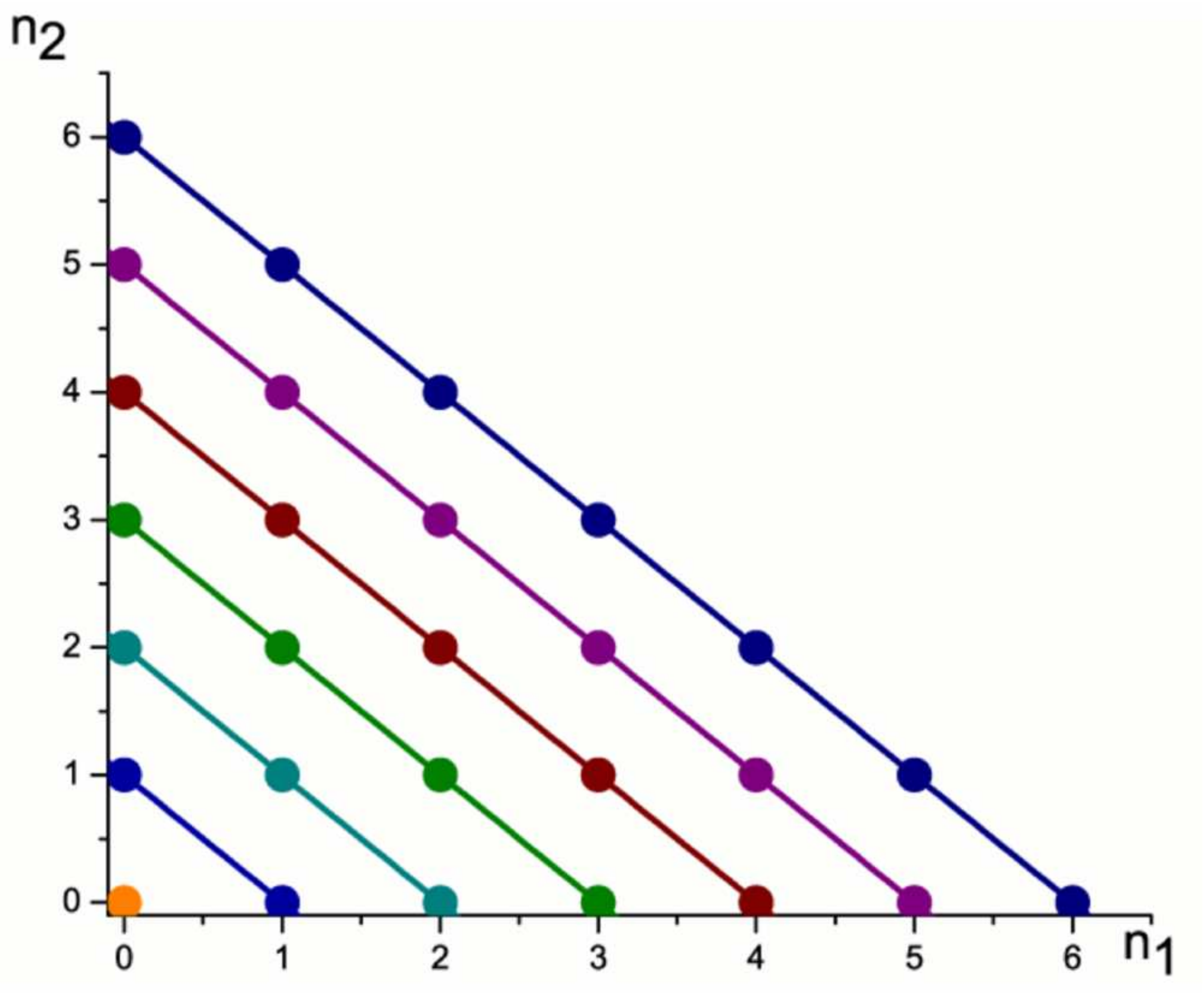}
  \caption{Estados excitados de un oscilador bidimensional. Los estados
  con energ\'ia menor o igual que $\Delta E$ conforman un tri\'angulo,
  cuya \'area viene dada por la parte derecha de la Ec. (\ref{ec5}).}
  \label{fig2}
\end{figure}

En primera aproximaci\'on, las frecuencias de peque\~nas oscilaciones
alrededor del m\'inimo de $U$ dan el espectro de excitaciones del
sistema. En vez de seguir el procedimiento de calcular la funci\'on de
onda, construir $U$, etc, nosotros tomaremos el espectro medido del
\'atomo y lo ajustaremos con dos par\'ametros: una frecuencia promedio,
$\omega$ y una dimensi\'on efectiva, $d$, que da cuenta del n\'umero de 
grados de libertad que participan de las oscilaciones.

\subsection{El oscilador en $d$ dimensiones}

Consideremos un oscilador de frecuencia $\omega$ en $d$ dimensiones. La
constante $\hbar$ se toma igual a uno. El n\'umero de estados con 
energ\'ia de excitaci\'on menor o igual a $\Delta E$ es:

\begin{equation}\label{ec5}
N_{estados}\approx \frac{(\Delta E/\omega)^d}{d!}.
\end{equation}

\noindent
En efecto, $\Delta E/\omega$ es el n\'umero m\'aximo de cuantos de los
estados inclu\'idos y $(\Delta E/\omega)^d/d!$ el volumen de la regi\'on
con energ\'ia menor o igual a $\Delta E$ en el espacio de los n\'umeros
cu\'anticos $n_\alpha$. En $d=2$, por ejemplo, la parte derecha de la 
Ec. (\ref{ec5}) da el \'area del tri\'angulo correspondiente en la Fig. 
\ref{fig2}.

\begin{figure}[ht]
  \includegraphics[width=.8\linewidth,angle=0]{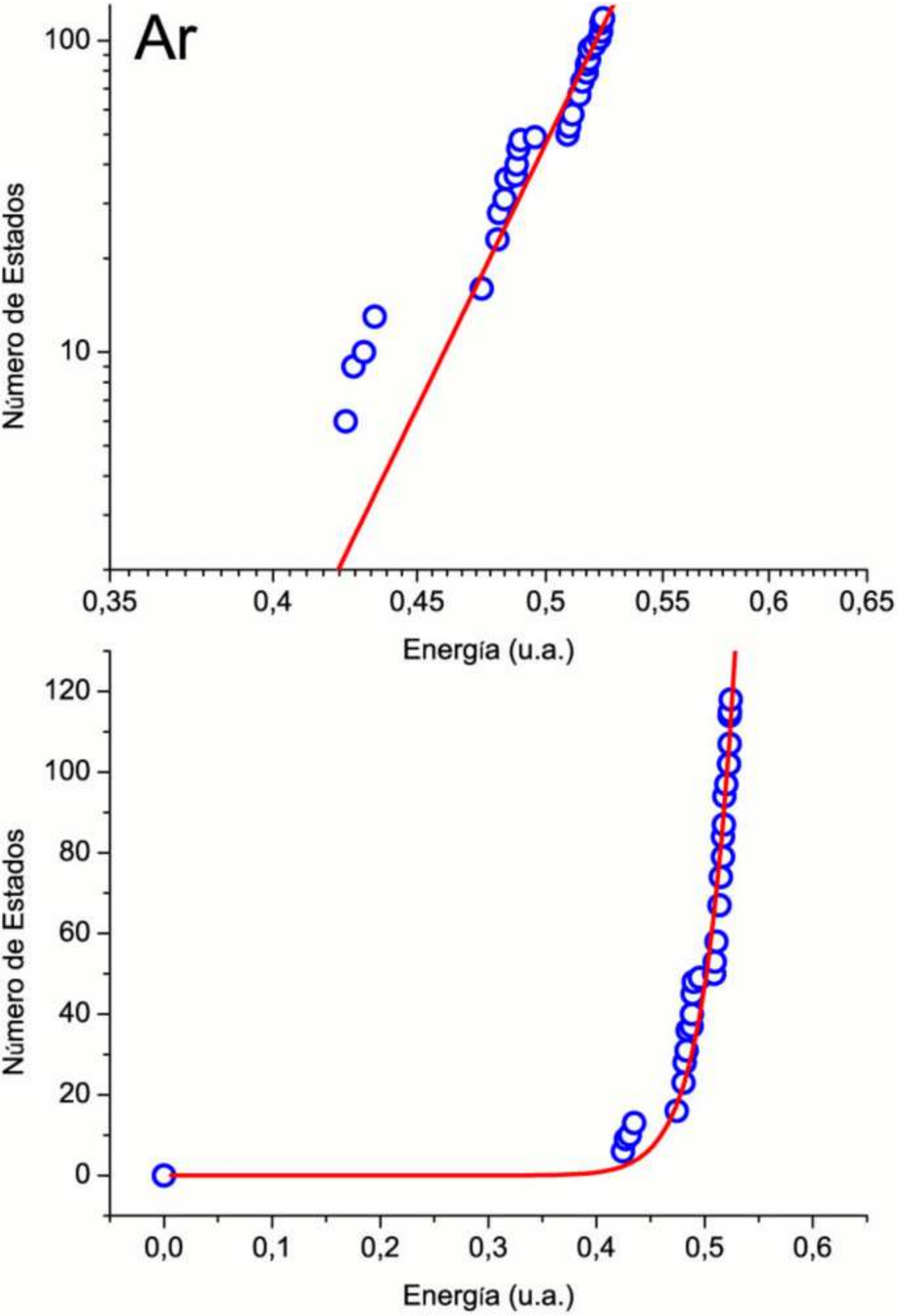}
  \caption{Espectro de excitaciones del Ar. Arriba: ajuste inicial 
  del parámetro $d$ con las excitaciones m\'as energ\'eticas. Debajo: 
  ajuste simultáneo de los parámetros $d$ y $\omega$ para todo el
  espectro.}
  \label{fig3}
\end{figure}

Al ajustar un espectro at\'omico permitiremos al par\'ametro $d$ tomar
valores no enteros, lo cual puede entenderse como un valor promedio
entre las distintas $d$ que aparecen a distintas energ\'ias de
excitaci\'on. Por eso en el denominador de la Ec. (\ref{ec5})
sustituiremos $d!$ por $\Gamma(d+1)$.

\section{Los espectros de excitaciones de \'atomos}

Se utilizaron los espectros de niveles de energ\'ia ofrecidos por el 
NIST \cite{NIST} para los \'atomos de la Tabla Peri\'odica con $2\le
Z\le 36$, desde el He hasta el Kr. Para cada \'atomo tomamos
los primeros 30 niveles con el objetivo de ajustar el espectro seg\'un
la f\'ormula (\ref{ec5}). En el n\'umero de estados con energ\'ia menor
o igual a $\Delta E$ se incluye expl\'icitamente la multiplicidad de
cada estado debida al momento total $J$. 

El n\'umero 30 de niveles inclu\'idos es completamente arbitrario y, en
\'atomos distintos, puede corresponder a diferentes proporciones de 
excitaciones intracapas e intercapas, las cuales exhiben diferentes
comportamientos en la energ\'ia de excitaci\'on \cite{mio2}. Sin
embargo, la experiencia nos sugiere \cite{mio3} fijar el n\'umero de 
niveles al comparar sistemas cu\'anticos con distintos n\'umeros de
part\'iculas.

\begin{table}[ht]
\begin{tabular}{|ccccc|}
\hline
 \text{Elemento} & Z & d & w & \text{Pot. Ioniz. (u.a.)} \\
 \hline
 \text{He} & 2 & 28.19 & 0.066 & 0.9040 \\
 \text{Li} & 3 & 8.21 & 0.027 & 0.1982 \\
 \text{Be} & 4 & 7.33 & 0.048 & 0.3428 \\
 \text{B} & 5 & 6.25 & 0.044 & 0.3051 \\
 \text{C} & 6 & 3.94 & 0.048 & 0.4140 \\
 \text{N} & 7 & 3.96 & 0.062 & 0.5344 \\
 \text{O} & 8 & 10.32 & 0.062 & 0.5007 \\
 \text{F} & 9 & 9.36 & 0.082 & 0.6406 \\
 \text{Ne} & 10 & 13.81 & 0.089 & 0.7929 \\
 \text{Na} & 11 & 9.81 & 0.024 & 0.1890 \\
 \text{Mg} & 12 & 8.02 & 0.038 & 0.2811 \\
 \text{Al} & 13 & 6.90 & 0.030 & 0.2201 \\
 \text{Si} & 14 & 8.38 & 0.034 & 0.2997 \\
 \text{P} & 15 & 8.90 & 0.045 & 0.3856 \\
 \text{S} & 16 & 10.15 & 0.045 & 0.3809 \\
 \text{Cl} & 17 & 11.61 & 0.053 & 0.4768 \\
 \text{Ar} & 18 & 18.56 & 0.052 & 0.5794 \\
 \text{K} & 19 & 8.46 & 0.021 & 0.1596 \\
 \text{Ca} & 20 & 5.71 & 0.027 & 0.2248 \\
 \text{Sc} & 21 & 3.43 & 0.010 & 0.2413 \\
 \text{Ti} & 22 & 1.73 & 0.003 & 0.2511 \\
 \text{V} & 23 & 5.46 & 0.010 & 0.2481 \\
 \text{Cr} & 24 & 6.57 & 0.016 & 0.2488 \\
 \text{Mn} & 25 & 2.88 & 0.013 & 0.2733 \\
 \text{Fe} & 26 & 6.08 & 0.013 & 0.2906 \\
 \text{Co} & 27 & 1.61 & 0.004 & 0.2898 \\
 \text{Ni} & 28 & 7.60 & 0.019 & 0.2809 \\
 \text{Cu} & 29 & 3.64 & 0.030 & 0.2841 \\
 \text{Zn} & 30 & 10.74 & 0.043 & 0.3454 \\
 \text{Ga} & 31 & 8.08 & 0.030 & 0.2206 \\
 \text{Ge} & 32 & 9.37 & 0.033 & 0.2904 \\
 \text{As} & 33 & 8.68 & 0.042 & 0.3599 \\
 \text{Se} & 34 & 7.13 & 0.046 & 0.3586 \\
 \text{Br} & 35 & 14.11 & 0.043 & 0.4344 \\
 \text{Kr} & 36 & 12.25 & 0.059 & 0.5148\\
 \hline
\end{tabular}
\caption{Par\'ametros ajustados $d$ y $\omega$ y potenciales de 
ionizaci\'on medidos para los \'atomos con $2\le Z\le 32$.}
\label{tab1}
\end{table}

En la Fig. \ref{fig3} se muestra, en el ejemplo del Ar, el procedimiento 
seguido para determinar los par\'ametros $d$ y $\omega$. Recordar que
nos limitamos a los primeros 30 niveles de energ\'ia. Utilizando
inicialmente escala doble logar\'itmica y fij\'andonos s\'olo en las 
excitaciones de mayor energ\'ia hallamos un valor para $d$. Ese valor 
inicial es usado como punto de partida para el ajuste de todo el 
espectro con la f\'ormula (\ref{ec5}). Los par\'ametros obtenidos se 
muestran en la Tabla \ref{tab1}.

\begin{figure}[ht]
  \includegraphics[width=\linewidth,angle=0]{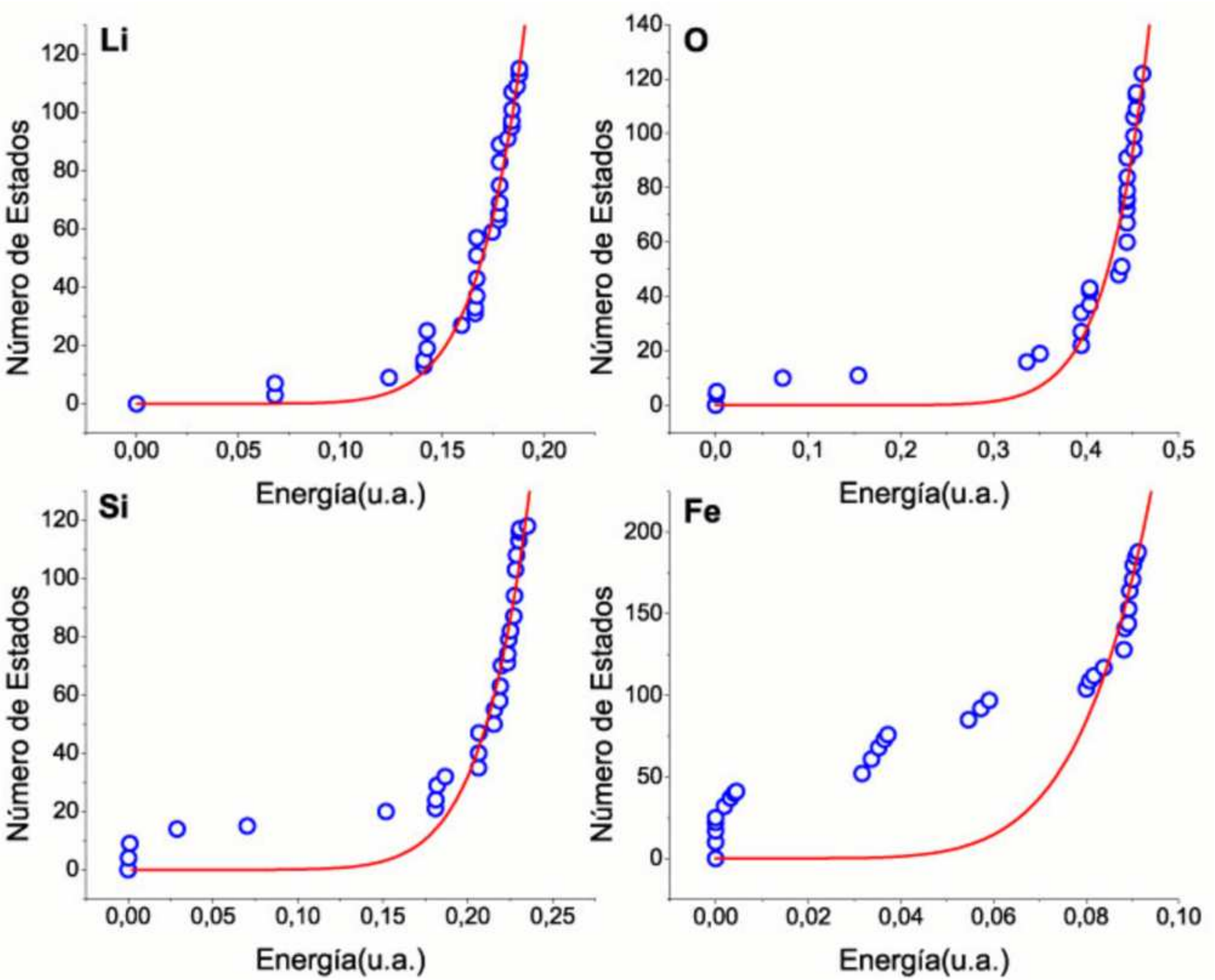}
  \caption{Espectros de excitaciones de bajas energías y ajustes seg\'un
  la Ec. (\ref{ec5}) para diferentes \'atomos.}
   \label{fig4}
\end{figure}

En la Fig. \ref{fig4} mostramos que la calidad del ajuste var\'ia para
diferentes \'atomos. En \'atomos con capas abiertas, las excitaciones
intracapas  no son muy bien descritas por la ley (\ref{ec5}). Si nos
hubi\'eramos extendido hasta m\'as altas energ\'ias tomando, por ejemplo, 100
niveles en vez de 30, los ajustes hubieran parecido cualitativamente mucho
mejores. 

\begin{figure}[ht]
  \includegraphics[width=.8\linewidth,angle=0]{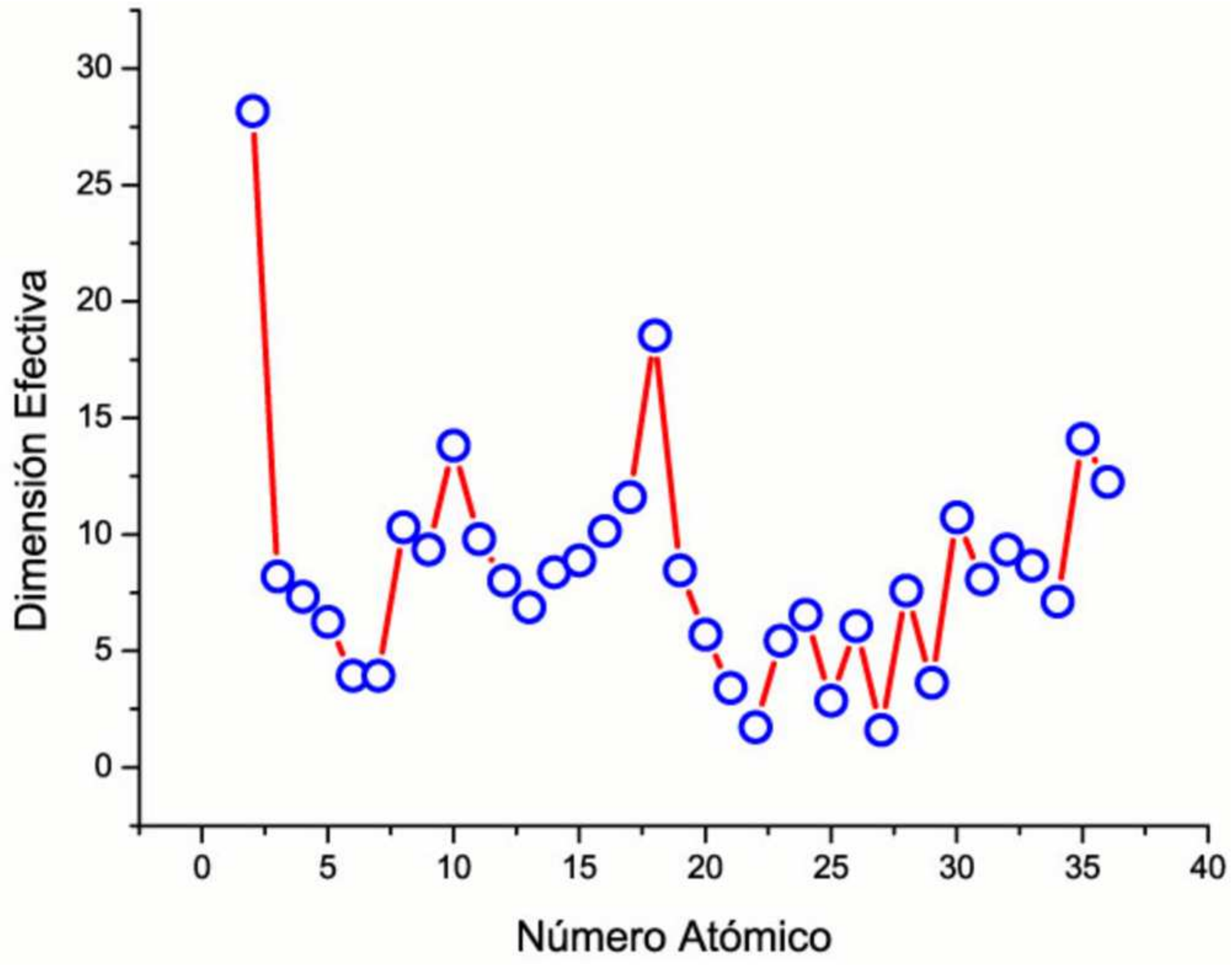}
  \caption{Relación entre la dimensión efectiva $d$ y el número atómico
   $Z$.}\label{fig5}
\end{figure}

La Fig. \ref{fig5} ilustra la dependencia del par\'ametro $d$ con el n\'umero
at\'omico $Z$, evidenciando claros efectos de llenado de capas. Los \'atomos 
con capas llenas tienen valores de $d$ m\'as altos. Si interpretamos $d/3$ como
el n\'umero efectivo de electrones que participan en las oscilaciones, vemos
que para el He este n\'umero es 9 (un valor no f\'isico), pero para el Ne es
casi 5, para el Ar es 6 y para el Kr es 4. Todos estos son n\'umeros
razonables que indican que en las excitaciones de m\'as bajas energ\'ias
participan colectivamente los electrones de la \'ultima capa.

\begin{figure}[ht]
  \includegraphics[width=.8\linewidth,angle=0]{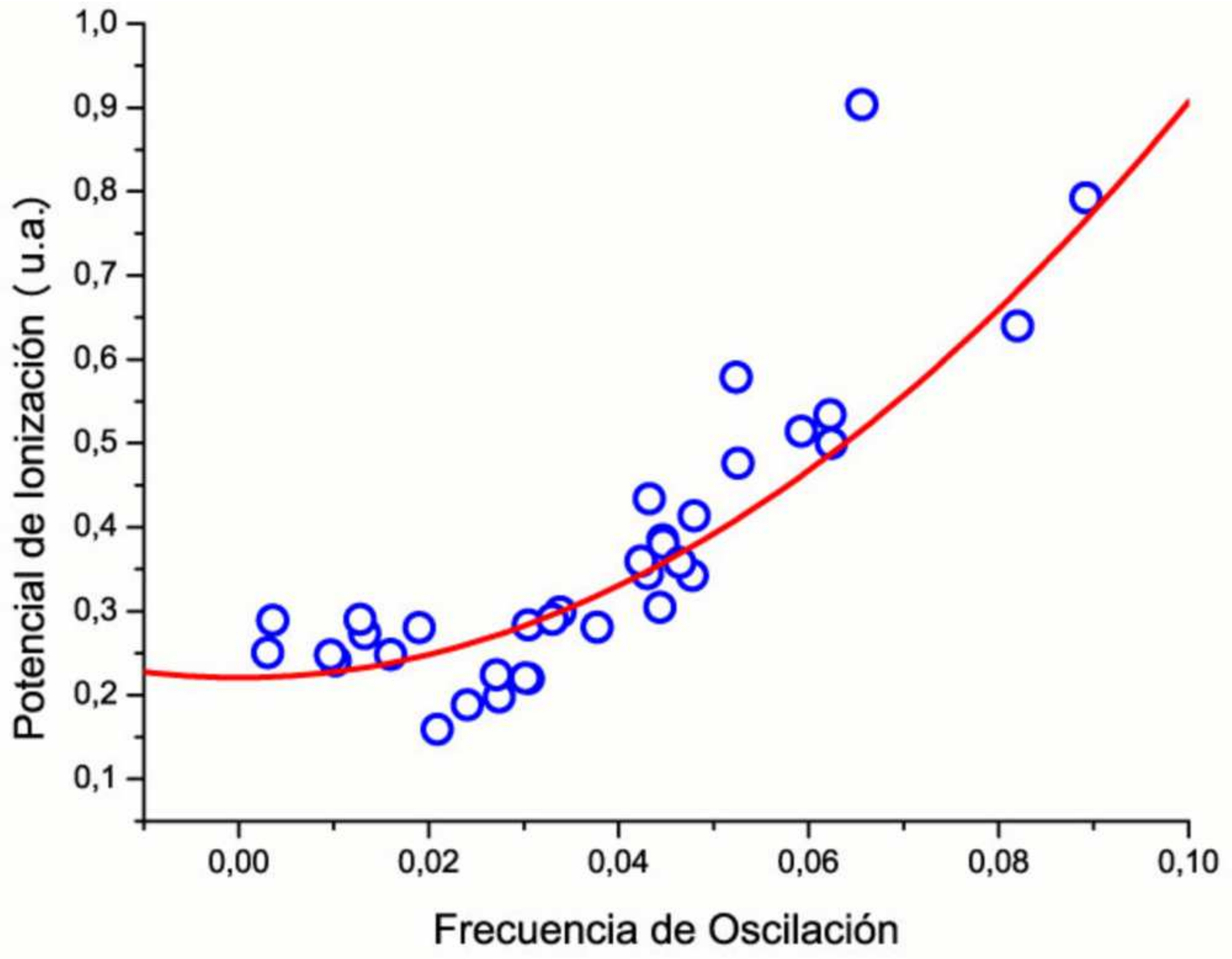}
  \caption{Relación entre el Potencial de Ionización y la frecuencia 
  de oscilación $\omega$. La línea roja es s\'olo una gu\'ia visual.}
  \label{fig6}
\end{figure}

Por otro lado, el par\'ametro $\omega$ es la curvatura efectiva del
superpotencial alrededor del m\'inimo $U=0$. Es m\'as dif\'icil relacionar este
par\'ametro con alguna magnitud at\'omica. En la Fig. \ref{fig6} comparamos el
potencial de ionizaci\'on del \'atomo con el valor ajustado de $\omega$. Estas
magnitudes deben estar relacionadas, aunque el potencial de ionizaci\'on
corresponde al tope del espectro discreto, es decir a m\'as altas energ\'ias de
excitaci\'on. La curva cont\'inua en la Fig. es s\'olo una gu\'ia visual, 
pero muestra que existe cierta correlaci\'on entre las dos magnitudes. 

\section{Conclusiones}

El presente trabajo se ha nutrido de varios ingredientes. Por un lado,
utilizando la denominada Mec\'anica Cu\'antica Supersim\'etrica, argumentamos
que las excitaciones de m\'as bajas energ\'ias de cualquier sistema cu\'antico
corresponden a osciladores arm\'onicos. Y planteamos un modelo sencillo para
el n\'umero de estados excitados con energ\'ia de excitaci\'on menor que
$\Delta E$. Este modelo tiene dos par\'ametros: el n\'umero de grados de
libertad, $d$ y la frecuencia promedio de los osciladores, $\omega$.

Por otro lado, utilizando los datos experimentales de los espectros at\'omicos
ofrecidos por el NIST, determinamos los par\'ametros $d$ y $\omega$ para todos
los \'atomos con $2\le Z\le 36$. Para ello se utilizaron los primeros 30
niveles excitados de cada \'atomo. En realidad, si hubi\'eramos
utilizado m\'as
niveles, por ejemplo 100, la calidad del ajuste hubiera sido muy superior. No
lo hicimos porque no tenemos respuesta a la pregunta de por qu\'e 
aparentemente no hay efectos anarm\'onicos a energ\'ias m\'as altas. Es
decir, por qu\'e los \'atomos se comportan casi como osciladores
perfectos. Las  variaciones del par\'ametro $d$ se relacionan con el 
llenado de las capas at\'omicas y ``muestran'' el car\'acter colectivo 
de las excitaciones en los \'atomos con capas llenas. El par\'ametro 
$\omega$, por su parte, evidencia cierta correlaci\'on con el potencial 
de ionizaci\'on del \'atomo.

Los ajustes de los espectros at\'omicos con el modelo simplificado de
osciladores, dado por la Ec. (\ref{ec5}), son cualitativamente buenos. Lo cual
trae de vuelta, como muchas veces sucede en la F\'isica y en otras ciencias,
a un viejo modelo, descartado en su momento, el modelo de Thomson para los 
\'atomos, donde el potencial de confinamiento era arm\'onico.

{\bf Agradecimientos}

Los autores agradecen a los participantes del Seminario de F\'isica
Te\'orica del ICIMAF por las valiosas discusiones sobre el tema.
El trabajo recibi\'o apoyo financiero del Programa Nacional de Ciencias 
B\'asicas (Cuba) y de la Red Caribe\~na de Mec\'anica Cu\'antica,
Part\'iculas y Campos (ICTP, Trieste).

\end{document}